\documentclass[prl,aps,twocolumn,showpacs,floatfix,longbibliography]{revtex4-2}

\usepackage{mathrsfs}
\usepackage{rotating}
\usepackage{color}
\usepackage{graphicx}
\usepackage{dcolumn}
\usepackage{bm}
\usepackage{hyperref}
\usepackage[latin1]{inputenc}
\usepackage{pstricks}
\usepackage{amsmath,amssymb}
\usepackage{version}
\usepackage{epstopdf}
\DeclareGraphicsExtensions{.eps,.ps}
\usepackage{ulem}
\usepackage{float}
\usepackage{booktabs}
\usepackage{subfigure}
\usepackage{setspace}
\usepackage{soul}

\begin{document}
\title{Ultrafast Ionization Dynamics Encoded in a Photoelectron Spin Torus}

\author{Xiaodan Mao}
\affiliation{State Key Laboratory of Dark Matter Physics, Key Laboratory for Laser Plasmas (Ministry of Education) and School of Physics and Astronomy, Collaborative Innovation Center for IFSA (CICIFSA), Shanghai Jiao Tong University, Shanghai 200240, China}
\author{Feng He}
\affiliation{State Key Laboratory of Dark Matter Physics, Key Laboratory for Laser Plasmas (Ministry of Education) and School of Physics and Astronomy, Collaborative Innovation Center for IFSA (CICIFSA), Shanghai Jiao Tong University, Shanghai 200240, China}
\author{Pei-Lun He}
\email{peilunhe@sjtu.edu.cn}
\affiliation{State Key Laboratory of Dark Matter Physics, Key Laboratory for Laser Plasmas (Ministry of Education) and School of Physics and Astronomy, Collaborative Innovation Center for IFSA (CICIFSA), Shanghai Jiao Tong University, Shanghai 200240, China}

\date{\today}
\begin{abstract}
We demonstrate that strong-field ionization of atoms in circularly polarized laser fields generates a photoelectron spin texture with toroidal topology in momentum space. 
Using time-dependent Schr\"odinger equation simulations, spin-resolved classical-trajectory Monte Carlo calculations, and an extended spin-resolved strong-field approximation including intermediate excitation pathways, we show that the rotation angle of this spin torus provides access to attosecond relative time delays associated with photoelectron wave packets released by tunneling from the counter-rotating and co-rotating \(p\)-orbital channels.
When intermediate-state dynamics become significant, the torus develops a clear splitting.
These results establish photoelectron spin textures as a complementary source of dynamical information beyond conventional momentum spectroscopy, and identify spin polarization as a robust internal degree of freedom for self-referenced attosecond metrology.
\end{abstract}

\maketitle

In strong-field physics, the photoelectron momentum distribution (PMD) is a central probe of ionization dynamics. A prominent example is the attoclock technique, which relates the angular offset of PMDs generated by elliptically polarized light to an ionization time delay~\cite{eckle2008attosecondDelay,eckle2008attosecond,pfeiffer2012attoclock}.
However, the extracted time delay is inherently ambiguous, as the same angular offset also contains Coulomb-induced deflections accumulated during continuum propagation~\cite{torlina2015interpreting,ni2016tunneling,PhysRevLett.119.023201,bray2018keldysh,sainadh2019attosecond,ge2024spatiotemporal}.
Disentangling these contributions remains a central challenge in attosecond metrology~\cite{PhysRevLett.129.203201}.

Electron spin offers a route around this ambiguity. For circularly polarized pulses, preferential ionization from counter-rotating \(p\) orbitals produces circular dichroism~\cite{barth2011nonadiabatic,herath2012strong,barth2013nonadiabatic,eckart2018ultrafast,dubois2024energy,mao2025modification}, which, together with bound-state spin--orbit coupling (SOC), gives rise to substantial longitudinal spin polarization~\cite{barth2013spin,hartung2016electron,trabert2018spin,milovsevic2016possibility,liu2018energy,liu2018deformation,ruberti2024bell,PhysRevLett.126.054801,zhang2025spin}. Continuum SOC is relativistically weak under typical strong-field ionization conditions~\cite{hu2023effect,ivanov2026effect}, although it can become relevant near the Cooper minimum in single-photon ionization~\cite{fano1969spin,lambropoulos1973spin}, when correlated ionic dynamics are resolved~\cite{kubel2019spatiotemporal,PhysRevLett.129.173202,stewart2023attosecond,carlstrom2023control}, or when rescattering electrons enter the weakly relativistic regime~\cite{hilgner1967electron,maxwell2024relativistic}. Thus, for the nonrelativistic atomic strong-field ionization considered here, the Coulomb field strongly reshapes the outgoing momentum distribution while leaving the spin polarization along each trajectory largely intact \cite{bargmann1959precession}. The resulting momentum-resolved spin polarization, known as the photoelectron spin texture (PST), therefore provides a complementary dynamical observable that, when combined with the PMD, forms a self-referenced probe of strong-field ionization dynamics.

Recent work has shown that PSTs can exhibit vortex structures under linearly polarized driving~\cite{he2025photoelectron}, suggesting that topologically nontrivial PSTs exist more generally. This possibility is particularly intriguing in view of the broader importance of polarization textures in condensed-matter and optical systems, where topological robustness against smooth deformations underpins a wide range of transport \cite{hsieh2009observation,lv2015observation}, magnetic \cite{muhlbauer2009skyrmion,yu2010real,fert2013skyrmions,yu2018transformation}, and optical phenomena \cite{tsesses2018optical,du2019deep,dai2020plasmonic,shen2024optical,rao2025meron,ornelas2024non,ornelas2025topological}.

In this Letter, we investigate whether such topologically nontrivial PSTs arise in circularly polarized fields by combining time-dependent Schr\"odinger equation (TDSE) simulations,
spin-resolved classical-trajectory Monte Carlo (CTMC) calculations, and an extended spin-resolved strong-field approximation (eSFA) that includes intermediate excitation pathways. 
We show that strong-field ionization by a circularly polarized laser pulse produces a PST with
toroidal topology in three-dimensional momentum space, which we refer to as a spin torus. The rotation angle of this torus, referenced to the attoclock offset angle, provides a self-referenced probe of relative ionization time delays, on the order of attoseconds, between
wave packets emitted from the counter-rotating and co-rotating \(p\)-orbital channels. 
When intermediate-state dynamics become significant, the spin torus exhibits a clear splitting, highlighting its sensitivity to excitation-mediated processes.
These results establish PSTs as a new tool for attosecond metrology, with spin serving as a robust internal degree of freedom that is largely decoupled from continuum Coulomb distortions.

As a representative system, we consider the tunneling ionization of Xe driven by a circularly polarized laser field in the \(x\)--\(y\) plane,
\[
\boldsymbol{A}(t)=-A_0\big(\sin(\omega t+\Phi_{\rm CEP}),\cos(\omega t+\Phi_{\rm CEP}),0\big)f(t),
\]
where \(f(t)=\sin^2[\omega t/(2N)]\) is the pulse envelope, \(N\) the number of optical cycles, \(\omega\) the laser frequency, and \(\Phi_{\rm CEP}\) the carrier-envelope phase (CEP). Owing to the relatively low ionization potential of the valence shell, the photoelectron signal is dominated by ionization from the outer \(5p\) shell.
SOC splits this shell into the \(5p_{1/2}\) and \(5p_{3/2}\) sublevels; under the present conditions, the ionization signal is dominated by the \(5p_{3/2}\) channel.
Continuum SOC contributes only at relativistic order \(O(1/c^2)\) and is neglected at leading order \cite{bargmann1959precession,supp}, an approximation further justified by the strong suppression of electron recollision in long-wavelength circularly polarized fields \cite{PhysRevLett.105.083002,PhysRevLett.124.253203}. Under these conditions, the PST reduces to bilinear combinations of the \(p\)-orbital ionization amplitudes, as shown in Eq.~\eqref{spinangle}; further details are provided in the Supplemental Material~\cite{supp}.

\begin{figure}
\centering
\includegraphics[width=0.48\textwidth]{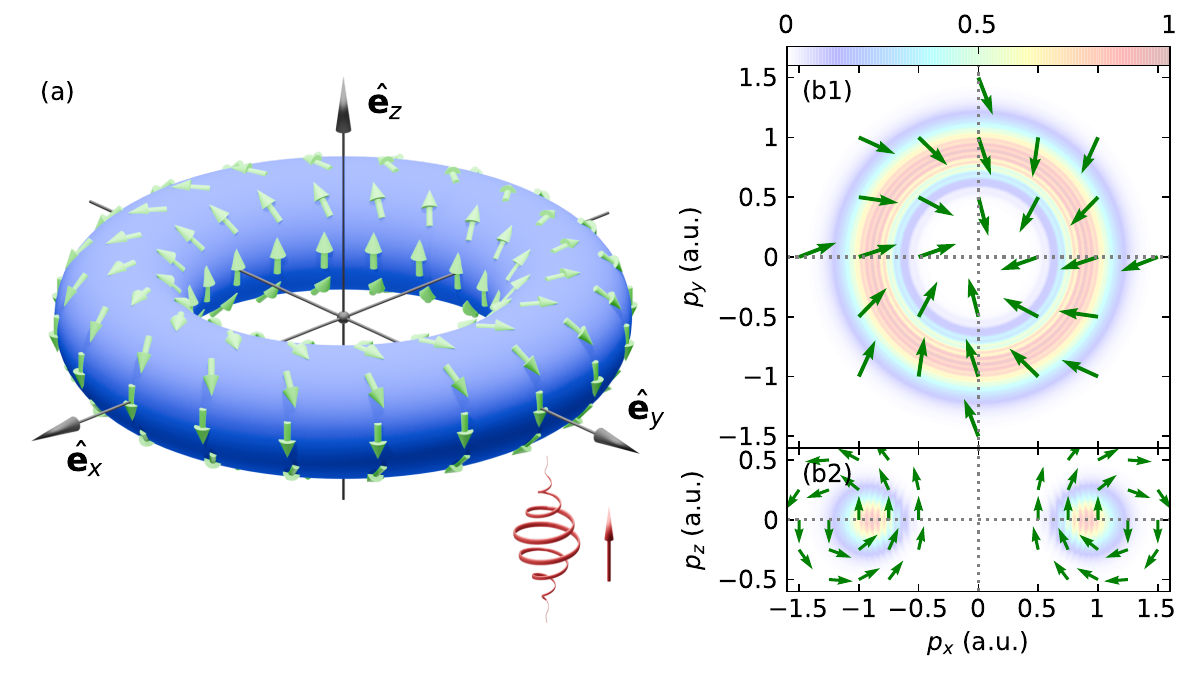}
\caption{Toroidal structure of the PST. 
(a) Three-dimensional visualization of the PST; green arrows indicate the spin-polarization direction, and the red curve marks the rotation direction of the laser electric field. (b1,b2) Spin polarization on two representative momentum-space cuts: (b1) the \(x\)--\(y\) plane at \(p_z=-0.02\) a.u. and (b2) the \(x\)--\(z\) plane at \(p_y=0\) a.u. The background colormap shows the corresponding PMD. Laser wavelength: 1030~nm; peak intensity: \(10^{14}\)~W/cm\(^2\); number of optical cycles: \(N=4\); averaged over the CEP.}
\label{FigPSTTorus}
\end{figure}

Figure~\ref{FigPSTTorus} shows the PST obtained from TDSE simulations using the Tong--Lin effective potential \cite{zhang2023qpc,tong2005empirical}. As shown in Fig.~\ref{FigPSTTorus}(a), the PMD forms a toroidal support in momentum space on which the PST is defined. Representative cuts of the PST are displayed in Fig.~\ref{FigPSTTorus}(b1,b2). 
At \(p_z=0\), the spin polarization points along the laser-propagation direction, with vanishing transverse components. Because the centroid of the PMD does not coincide with the spin nodal point, defined by \(\boldsymbol{\zeta}(\boldsymbol{p})=0\), the distribution acquires a net longitudinal polarization \cite{barth2013spin}. For finite \(p_z\), transverse components emerge: the polarization winds clockwise for \(p_z<0\) and counterclockwise for \(p_z>0\) [Fig.~\ref{FigPSTTorus}(b1)]. In the \(x\)--\(z\) plane [Fig.~\ref{FigPSTTorus}(b2)], the projected spin texture forms a vortex centered at the nodal point, and the associated planar topology is characterized by a unit winding number defined through the spin polarization along a closed momentum-space loop encircling the node~\cite{supp}. 
Taken together, these features define a toroidal spin texture in momentum space, which can be probed experimentally by combining Mott polarimetry with ultrafast spectroscopy \cite{hartung2016electron,trabert2018spin}. A full spin-tomographic reconstruction could be achieved using a Radon-transform-based approach~\cite{supp}.

To understand the origin of the toroidal structure, we exploit the rotational invariance of the atom--laser interaction in the \(x\)--\(y\) plane. In the spirit of the Coulomb-corrected SFA \cite{yan2010low,shvetsov2016semiclassical}, we parametrize the photoelectron wave function as
\begin{equation}
\left\{
\begin{aligned}
\chi^{(+)} &= \mathfrak{A}_+\, e^{i(\phi_{\boldsymbol{p}}+\Phi_c)} e^{i S_c} e^{i\delta^{(+)}}, \\
\chi^{(0)} &= \mathfrak{A}_0\, e^{i S_c} e^{i\delta^{(0)}}, \\
\chi^{(-)} &= \mathfrak{A}_-\, e^{-i(\phi_{\boldsymbol{p}}+\Phi_c)} e^{i S_c} e^{i\delta^{(-)}}.
\end{aligned}
\right.
\label{Generalamplitude}
\end{equation}
Here, \(\chi^{(+)}\), \(\chi^{(0)}\), and \(\chi^{(-)}\) correspond to the initial orbital magnetic quantum numbers \(m_l=1\), \(0\), and \(-1\), respectively, and the real coefficients \(\mathfrak{A}_{m_l}\) characterize the ionization strengths of the corresponding channels. 
The \(\phi_{\boldsymbol{p}}\)-dependent phase factor reflects the orbital angular momentum of the initial state, while \(\Phi_c\) denotes the Coulomb-induced angular offset in momentum space relative to the Coulomb-free case.
The semiclassical action \(S_c\) is accumulated during continuum propagation \cite{li2014classical}, and \(\delta^{(m_l)}\) represents additional phase contributions beyond the semiclassical ionization dynamics. All quantities depend on the radial momentum \(p_r=\sqrt{p_x^2+p_y^2}\) and the longitudinal momentum \(p_z\), although this dependence is suppressed for brevity. Within the standard SFA \cite{keldysh1965ionization,faisal1973multiple,reiss1980effect}, both \(\Phi_c\) and \(\delta^{(m_l)}\) vanish in the absence of the Coulomb interaction. For above-threshold ionization (ATI) \cite{milovsevic2006above} involving the absorption of \(n\) photons, \(S_c\) is proportional to \(-n\phi_{\boldsymbol{p}}\), reflecting energy and angular-momentum conservation \cite{supp}.

\begin{figure}
\centering
\includegraphics[width=0.48\textwidth]{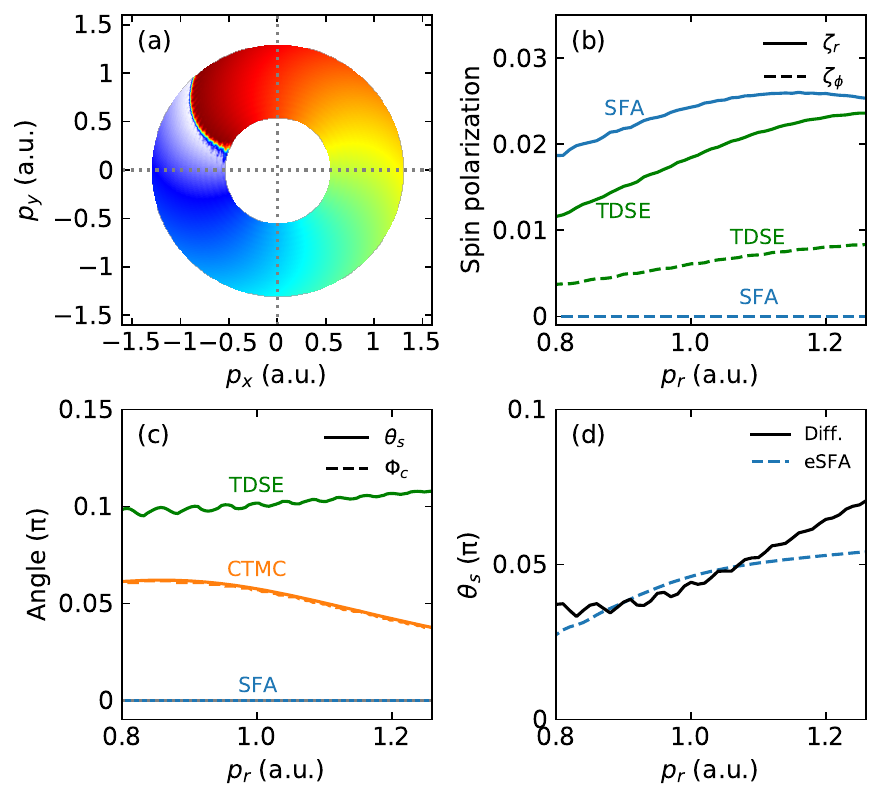}
\caption{Origin of the spin torus and its connection to attoclock dynamics. 
(a) Phase difference between the ionization amplitudes \(\chi^{(+)}\) and \(\chi^{(0)}\). 
(b) Radial \((\zeta_r)\) and azimuthal \((\zeta_\phi)\) spin-polarization components. The SFA result is purely radial, whereas the TDSE result exhibits a finite azimuthal component. 
(c) Spin-rotation angle \(\theta_s\) and Coulomb-induced angular shift \(\Phi_c\) as functions of radial momentum \(p_r\), obtained from TDSE and CTMC calculations. 
(d) Difference in \(\theta_s\) between the TDSE and CTMC results, together with the corresponding eSFA contribution. 
Here, \(\boldsymbol{e}_r=(\cos\phi_{\boldsymbol{p}},\sin\phi_{\boldsymbol{p}},0)\) and \(\boldsymbol{e}_\phi=(-\sin\phi_{\boldsymbol{p}},\cos\phi_{\boldsymbol{p}},0)\) are the radial and azimuthal unit vectors, respectively. 
Laser parameters are the same as those in Fig.~\ref{FigPSTTorus}.}
\label{FigSpinAngle}
\end{figure}

This parametrization exposes relative phase singularities between the ionization channels [Fig.~\ref{FigSpinAngle}(a)], analogous to optical vortices carrying orbital angular momentum \cite{allen1992orbital}, and thus reveals the origin of the spin vortex and its planar winding. Since this structure follows from angular-momentum conservation in the absorption of circularly polarized photons, it is intrinsically robust. Using Eq.~\eqref{Generalamplitude}, the PST reads
\begin{equation}
\left\{
\begin{aligned}
\zeta_0
&=\frac{4}{3}\left( \mathfrak{A}_+^2 + \mathfrak{A}_-^2 + \mathfrak{A}_0^2 \right), \\
\zeta_r &=
\frac{2\sqrt{2}}{3\,\zeta_0}\,\mathfrak{A}_0
\left[
   \mathfrak{A}_+ \cos\!\left(\theta^+\right)
 +
   \mathfrak{A}_- \cos\!\left(\theta^-\right)
\right], \\[4pt]
\zeta_\phi &=
\frac{2\sqrt{2}}{3\,\zeta_0}\,\mathfrak{A}_0
\left[
   \mathfrak{A}_+ \sin\!\left(\theta^+\right)
 +
   \mathfrak{A}_- \sin\!\left(\theta^-\right)
\right],\\
\zeta_z
&=\frac{2}{3\zeta_0}
\left( \mathfrak{A}_+^2 - \mathfrak{A}_-^2 \right),
\end{aligned}
\right.
\label{spinangle}
\end{equation}
where \(\zeta_0\) gives the PMD, and \(\zeta_r\), \(\zeta_\phi\), and \(\zeta_z\) are the radial, azimuthal, and longitudinal components of the spin polarization, respectively. 
The relative ionization phase between a channel pair \((m_i,m_j)\) is defined as
\(\Delta\delta^{(m_i,m_j)} \equiv \delta^{(m_i)}-\delta^{(m_j)}\).
Accordingly, the relative phases for the \((+,0)\) and \((0,-)\) channel pairs are defined as
\(\theta^+ \equiv \Delta\delta^{(+,0)}+\Phi_c\) and
\(\theta^- \equiv \Delta\delta^{(0,-)}+\Phi_c\),
both of which are manifestly gauge invariant.

Within the plain SFA, the azimuthal spin-polarization component
\(\zeta_\phi\) vanishes, so the spin polarization is purely radial
[Fig.~\ref{FigSpinAngle}(b)], providing a natural reference
direction. For \(p_z \neq 0\), the spin polarization acquires a
transverse component, and the deviation from the radial direction
defines the spin-rotation angle,
\begin{equation}
\theta_s \equiv \arg(\zeta_r + i\zeta_\phi),
\end{equation}
which depends only weakly on \(p_{z}\) \cite{supp}.

The CTMC simulations in Fig.~\ref{FigSpinAngle}(c) show that
classical Coulomb drift in the continuum induces a finite spin
rotation, in good agreement with the momentum offset angle. This
agreement is not accidental and admits a simple physical
interpretation: at the instant of ionization, the spin polarization
forms a vortex with respect to the instantaneous laser electric field
\cite{he2025photoelectron}. 
The subsequent laser and Coulomb forces then deflect the electron momentum without appreciably altering the spin polarization along the trajectory; the corresponding spin rotation due to the continuum SOC is only of order \(10^{-6}\,\mathrm{rad}\) for the parameters considered here \cite{supp}.
Thus, the agreement between the two angles reflects the angular
streaking dynamics in the continuum. Integrating over ionization times then yields the toroidal structure of the PST.

From Eqs.~\eqref{Generalamplitude} and \eqref{spinangle}, it follows that \(\theta_s=\Phi_c\) when \(\delta^{(m_l)}=0\). Any mismatch between the two angles, therefore, signals an additional nonclassical phase contribution associated with tunneling-ionization dynamics. In contrast to the CTMC results, the full TDSE results exhibit a clear mismatch between \(\theta_s\) and \(\Phi_c\) [Fig.~\ref{FigSpinAngle}(c)], indicating a contribution beyond classical Coulomb-induced deflection. The ATI structure of the photoelectron spectrum gives rise to small energy-dependent oscillatory modulations in \(\theta_s\). However, these modulations do not remove the finite mismatch between the two angles over the investigated range and therefore do not affect the central conclusion.

To identify the origin of this mismatch, we employ the eSFA, which explicitly incorporates transitions through intermediate bound states~\cite{supp}. Within the eSFA, the ionization amplitude is written as
\begin{equation}
\chi^{(m_l)} = -i \sum_n \int d\tau\, C_n^{(m_l)}(\tau)
\left\langle \boldsymbol{p}_V(\tau) \right| H_i(\tau) \left| n \right\rangle,
\end{equation}
where \(H_i(\tau)=\boldsymbol{x}\!\cdot\!\boldsymbol{E}(\tau)\) is the light--matter interaction Hamiltonian, \( | \boldsymbol{p}_V \rangle \) the Volkov state \cite{volkov1935class}, and \(C_n^{(m_l)}(\tau)\) the time-dependent projection amplitude onto the bound state \(\lvert n\rangle\), extracted from the TDSE simulation for the \(m_l\) channel~\cite{liu2022laser}.
As shown in Fig.~\ref{FigSpinAngle}(d), the eSFA accounts for a substantial fraction of the difference between the spin-rotation angle obtained from the TDSE simulations and the angular offset associated with continuum propagation in the CTMC calculations, indicating that the additional angular deflection arises predominantly from excitation-mediated ionization pathways. 
A residual discrepancy nevertheless remains after inclusion of the eSFA contribution. This remaining difference may reflect either higher-order contributions beyond the perturbative eSFA treatment or nontrivial Coulomb dynamics in the sub-barrier region \cite{klaiber2013under}.

\begin{figure}
\centering
\includegraphics[width=0.45\textwidth]{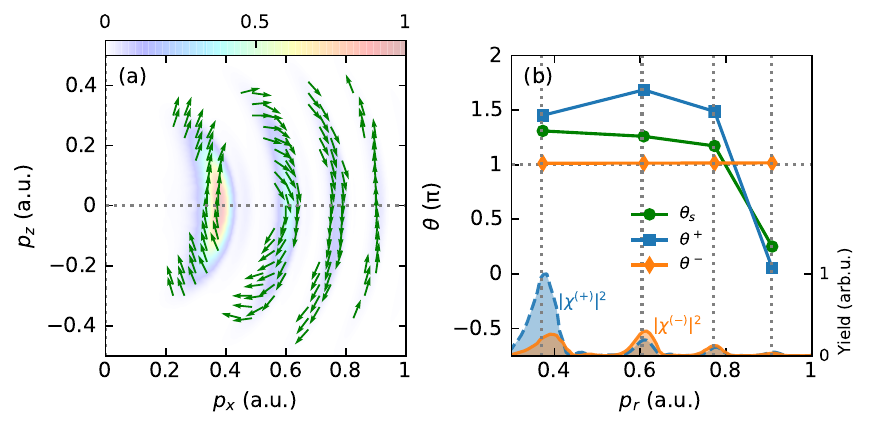}
\caption{
Splitting of the spin torus induced by intermediate excited-state dynamics. 
(a) Spin vortex in the \(x\)--\(z\) plane; green arrows indicate the local spin-polarization direction. 
(b) Relative phase angles \(\theta^{+}\) and \(\theta^{-}\), together with the spin-rotation angle \(\theta_s\), as functions of \(p_r\). The shaded regions denote the partial photoelectron yields \( |\chi^{(+)}|^2 \) and \( |\chi^{(-)}|^2 \), plotted on the right axis. Laser wavelength: 400~nm; intensity: \(I=10^{14}\)~W/cm\(^2\); number of optical cycles: \(N=10\); CEP: $\Phi_{\rm CEP}=0$.
}
\label{FigExcited}
\end{figure}

The sensitivity of the PST demonstrated above makes it an ideal probe for resolving intermediate excited states, whose presence is otherwise obscured in conventional PMDs \cite{PhysRevLett.116.123901}. The winding number associated with the azimuthal phase in the \(x\)--\(y\) plane is fixed by the magnetic quantum number and is therefore protected by angular-momentum conservation. 
When intermediate-state dynamics become significant, however, the spin texture on the poloidal cross-section undergoes a pronounced restructuring, manifested as a splitting of the spin torus, as shown in Fig.~\ref{FigExcited}(a), leading to a clear deviation from the SFA prediction \cite{supp}.
The underlying momentum-resolved phase analysis in Fig.~\ref{FigExcited}(b) reveals that \(\theta^-\) remains
close to \(\pi\) and is nearly independent of the radial momentum
\(p_r\) in the parameter regime considered here, so that the
contribution \(\mathfrak{A}_-\sin(\theta^-)\) to \(\zeta_\phi\) is
weak and acts only as an approximately momentum-independent
background. By contrast, \(\theta^+\) exhibits a strong \(p_r\)
dependence and therefore dominates the variation of \(\zeta_\phi\) and
thus of \(\theta_s\), driving the reversal of the spin rotation and,
hence, the splitting of the spin torus. 
In contrast to conventional circular dichroism \cite{eckart2018ultrafast}, the counter-rotating channel here exhibits slightly higher yields in the high-energy regime due to the excitations. 
The observed splitting thus provides a direct
fingerprint of the excitation-assisted pathway.

Beyond resolving intermediate excitation pathways, the PST also encodes the relative ionization phase \(\Delta\delta^{(m_i,m_j)}\) between different channels. Focusing on the electron dynamics in the polarization plane, we define the relative emission time delay as
\begin{equation}
\label{eq:td}
\Delta t_d^{(m_i,m_j)}
=
\frac{1}{p_r}\frac{\partial \Delta\delta^{(m_i,m_j)}}{\partial p_r}.
\end{equation}
This quantity measures the relative temporal shift between electron wave packets emitted through different ionization channels in the \(x\)--\(y\) plane, rather than an absolute ionization time delay associated with a single channel.

\begin{figure}
\centering
\includegraphics[width=0.48\textwidth]{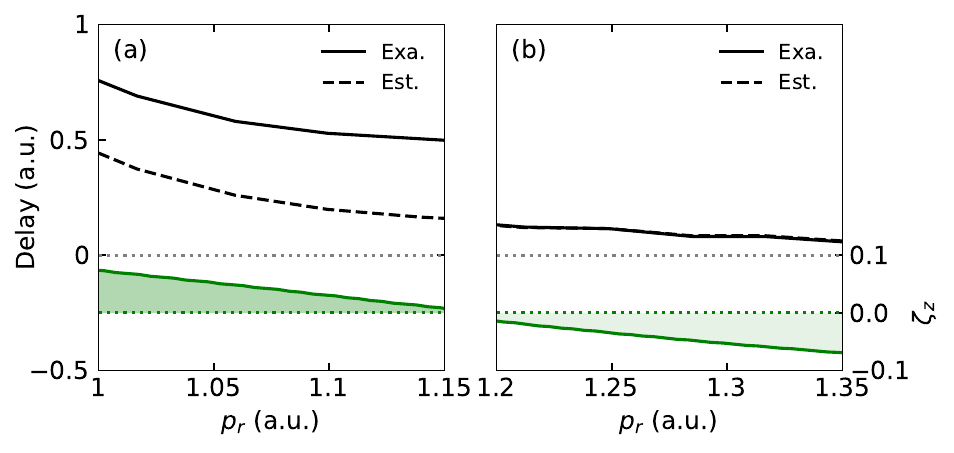}
\caption{Extraction of the relative emission time delay from the PST in (a) the low-energy region dominated by the counter-rotating channel, where the delay corresponds to the \((+,0)\) channels, and (b) the high-energy region dominated by the co-rotating channel, where the delay corresponds to the \((0,-)\) channels. 
Solid black curves show the exact results obtained directly from the TDSE wave functions, while dashed blue curves show the estimates from Eq.~\eqref{eq:td_est}. The shaded regions indicate the longitudinal spin polarization \(\zeta_z\) (right axis). Laser parameters are the same as those in Fig.~\ref{FigPSTTorus}, except that $\Phi_{\rm CEP}=0$.}
\label{FigBIC}
\end{figure}

Nevertheless, extracting the channel-resolved phases from Eq.~\eqref{spinangle} is, in general, underdetermined, since four independent relations constrain five unknown quantities. Resolving the missing information would normally require additional input from more sophisticated theoretical modeling or experimental techniques \cite{tan2018determination,han2019unifying,han2021complete,yu2022full,xie2024observation}. 
In the present case, however, circular dichroism \cite{barth2011nonadiabatic} provides a natural simplification by separating the spectrum into two asymptotic energy regimes: in the low-energy region, the counter-rotating channel dominates, so that the extracted signal is primarily sensitive to the phase difference \(\Delta\delta^{(+,0)}\), whereas in the high-energy region, the co-rotating channel prevails, and the signal mainly reflects \(\Delta\delta^{(0,-)}\).
Accordingly, the radial derivative of the spin-rotation angle reduces to
\begin{equation}
\label{eq:estimation}
\frac{\partial \theta_s}{\partial p_r}
\approx
\begin{cases}
\frac{\partial \theta^+}{\partial p_r}, & \text{for } \zeta_z > 0, \\
\frac{\partial \theta^-}{\partial p_r}, & \text{for } \zeta_z < 0.
\end{cases}
\end{equation}
Within the standard attoclock approximation, the Coulomb-induced angular shift satisfies \(\Phi_c \approx \theta_{\mathrm{atto}}\), where \(\theta_{\mathrm{atto}}\) is the angular offset of the PMD peak and can be extracted experimentally using a few-cycle circularly polarized pulse \cite{supp}. 
Equation~\eqref{eq:td} then becomes
\begin{equation}
\label{eq:td_est}
\Delta t_d
\approx \frac{1}{p_r}\left(
\frac{\partial \theta_s}{\partial p_r}
-
\frac{\partial \theta_{\mathrm{atto}}}{\partial p_r}
\right),
\end{equation}
where \(\Delta t_d\) reflects the relative emission-time delay between the \((+,0)\) channels in the low-energy regime and between the \((0,-)\) channels in the high-energy regime.
The approximation is expected to be most accurate in the corresponding asymptotic limits; away from these limits, residual competition between the channels may lead to deviations \cite{supp}.
As shown in Fig.~\ref{FigBIC}, the approximate curve captures the overall trend of the exact result in the low-energy region, with the main discrepancy appearing as an approximately uniform offset of about \(0.2\)~a.u.; at higher energies, the two curves become nearly indistinguishable.

Equation~\eqref{eq:td_est} therefore implies that the mismatch between the spin-rotation angle \(\theta_s\) and the attoclock offset angle \(\theta_{\mathrm{atto}}\) encodes a relative ionization time delay. 
In the present case, the extracted momentum-dependent relative delay is on the order of 10 attoseconds, with the co-rotating contribution emitted slightly earlier than the counter-rotating one.
Although this delay reflects the underlying tunneling-ionization dynamics, it should be interpreted as the relative emission-time delay between the two wave packets rather than as the ionization time defined at the tunnel exit \cite{liu2018deformation,PhysRevLett.129.203201}.
This relative delay is experimentally accessible through joint measurements of the PST and PMD, which together provide a self-referenced probe of tunneling dynamics. 
The mismatch between the spin-rotation angle and the attoclock offset persists under intensity variations, inclusion of the \(p_{1/2}\) channel, and other subleading corrections; see End Matter and Supplemental Material for details \cite{supp}. A spin-angle mismatch of order \(0.05\pi\) is already within reach of current experimental capabilities \cite{supp}.

In conclusion, we establish the PST as a robust observable for attosecond metrology, providing access to ultrafast ionization dynamics beyond conventional PMDs. By comparing the spin-rotation angle with the attoclock offset, one obtains a self-referenced measure of the relative delay between tunneling-emitted wave packets. Excitation-mediated ionization pathways leave clear fingerprints in the PST, most notably through the splitting of the spin torus. More generally, the extraction of ionization time delays from the PST extends naturally to the single-photon limit, where it provides access to the relative scattering phase and the Wigner--Smith time delay free from continuum--continuum coupling (see End Matter). 
Taken together, these results highlight the PST as a new measurable degree of freedom for probing ultrafast dynamics in atoms and molecules.

\textit{Acknowledgments}\textemdash
This work was supported by the National Natural Science Foundation of China (NSFC) under Grant Nos.~12574377 and 12450405. 
The computations were performed on the Siyuan-1 cluster supported by the Center for High Performance Computing at Shanghai Jiao Tong University. 
P.-L. H. acknowledges support from the Pujiang Program of the Shanghai Baiyulan Talent Plan (Grant No.~24PJA046), the Xiaomi Young Scholar Program, the Shanghai Jiao Tong University 2030 Initiative, and the Yangyang Development Fund.
X. M. acknowledges support from the Shanghai Super Postdoctoral Fellowship Program.
\bibliography{references.bib}

\clearpage
\section*{End Matter}

In this End Matter, we examine the robustness of the mismatch between the spin-rotation angle and the attoclock offset angle with respect to variations in laser intensity, and show how the relative emission-time delay can be extracted from the PST in single-photon ionization.

\subsection*{Laser intensity dependence}

The mismatch between the spin-rotation angle \(\theta_s\) and the attoclock offset angle \(\theta_{\mathrm{atto}}\) is related to a relative emission-time delay through Eq.~\eqref{eq:td_est}.
As discussed in the main text, this quantity should be interpreted as a relative emission-time delay between channel-resolved photoelectron wave packets, rather than as an absolute ionization time delay of a single channel.

\begin{figure}
\centering
\includegraphics[width=0.48\textwidth]{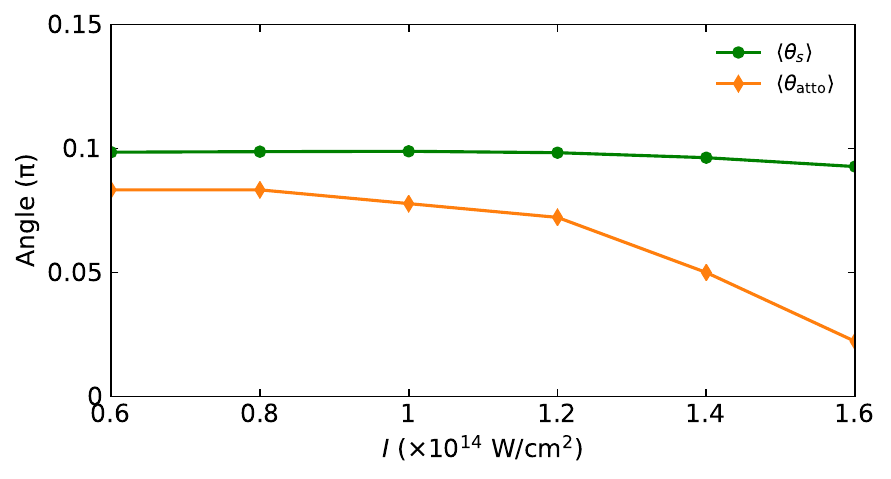}
\caption{Average spin-rotation angle \(\langle \theta_s \rangle\) and attoclock offset angle \(\langle \theta_{\mathrm{atto}} \rangle\) as functions of the peak laser intensity. A finite mismatch between the two angles persists throughout the investigated intensity range. 
Laser wavelength: 1030~nm; number of optical cycles: \(N=4\); CEP: $\Phi_{\rm CEP}=0$.}
\label{FigIntensity}
\end{figure}

Although the relative emission-time delay is, in general, \(p_r\)-dependent, a finite \(p_r\)-averaged mismatch already provides direct evidence that the delay does not vanish identically across the photoelectron distribution: if the underlying relative delay were trivial for all \(p_r\), the averaged mismatch would also vanish. 
Figure~\ref{FigIntensity} shows the \(p_r\)-averaged spin-rotation angle and attoclock offset angle as functions of laser intensity. In the present case, the average is dominated by counter-rotating electrons and therefore mainly reflects the behavior of the \((+,0)\) channel pair. The \(p_r\)-averaged angles thus provide experimentally convenient indicators of nontrivial tunneling-ionization dynamics.

The mismatch persists throughout the investigated intensity range, demonstrating that the averaged angular offset is robust against variations in the laser intensity. Its magnitude remains on the order of \(0.05\pi\), already within reach of current experimental capabilities \cite{supp}. 
The mismatch also persists when the \(p_{1/2}\) channel is included, while nondipole corrections, continuum spin-orbit coupling, and magnetic-field--spin coupling remain subleading under the present conditions; see Supplemental Material for details \cite{supp}. 
More generally, many-body and other higher-order effects may quantitatively shift the absolute values of \(\theta_s\) and \(\theta_{\mathrm{atto}}\); however, eliminating their mismatch would require fine-tuning and is therefore unlikely. 
The angular mismatch should therefore be regarded as a robust signature of the underlying relative emission-time delay.

\subsection*{Single-photon extraction of relative scattering phase and Wigner time delay}
To further demonstrate the capability of the PST to extract relative phase and time-delay information, we consider single-photon ionization, for which the underlying physics becomes analytically transparent and thus provides a stringent benchmark.
In this limit, dipole selection rules strongly constrain the contributing partial waves, allowing the mapping between the PST and the relative scattering phase to be derived explicitly and compared directly with the corresponding phase obtained from the exact scattering state.

In contrast to pump--probe techniques such as RABBITT \cite{doi:10.1126/science.aao7043,li2025photoionization,han2023attosecond} and attosecond streaking \cite{schultze2010delay,biswas2020probing}, where the measured delay generally contains a probe-induced continuum--continuum contribution that must be disentangled from the intrinsic Wigner delay, the relative scattering phase is encoded here directly in the transverse components of the PST. In the present implementation, the attosecond pulse train provides a series of photon energies that drive single-photon ionization without the need for an additional probing field that induces continuum--continuum transitions.

\begin{figure}
\centering
\includegraphics[width=0.48\textwidth]{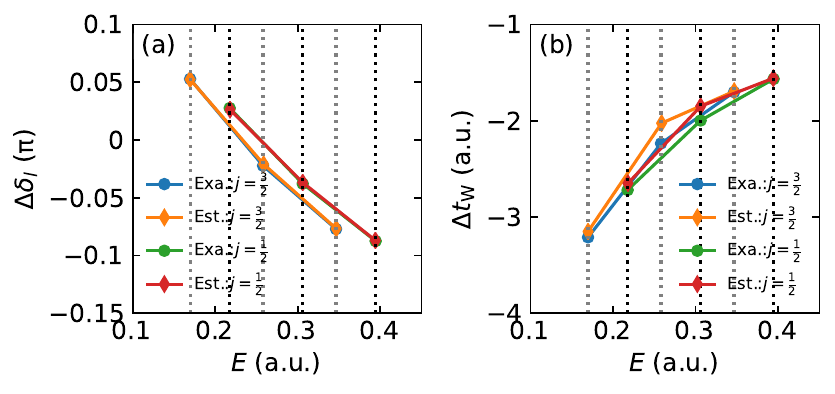}
\caption{Relative scattering phase and Wigner time-delay difference inferred from the PST in single-photon ionization as functions of the photoelectron energy. (a) Relative scattering phase \(\Delta\delta_l=\delta_0-\delta_2\) and (b) the corresponding Wigner time-delay difference \(\Delta t_W\) for the \(j=\tfrac{3}{2}\) and \(j=\tfrac{1}{2}\) channels. 
The fundamental wavelength is 1030 nm. The circularly polarized attosecond pulse train has a duration of \(N=5\) cycles of the fundamental driving field, a peak intensity of \(10^{12}\,\mathrm{W/cm^2}\), and a central harmonic order of 17.}
\label{FigSPI}
\end{figure}

The extraction of the relative phase, and hence of the time delay, is particularly transparent because the dipole selection rules allow only the outgoing partial waves with \(l=0\) and \(l=2\) for photoionization from the \(5p\) shell \cite{cherepkov1979spin}. The corresponding PST reads
\begin{equation}
\label{zetasingle}
\left\{
\begin{aligned}
 \zeta_0
=& \frac{1}{6\pi} \Bigg[
 F_0^2(p)
+ \frac{7-3\cos 2\theta}{4} F_2^2(p) \\
& \quad+ \frac{1+3\cos 2\theta}{2} F_0(p)F_2(p)
\cos\!\left(\delta_0-\delta_2\right)
\Bigg], \\
 \zeta_r
=& \frac{1}{\zeta_0}\frac{\sin 2\theta}{8\pi}\left[
F_2^2(p)
+\cos\!\left(\delta_0-\delta_2\right)F_0(p)F_2(p)\right], \\
 \zeta_\phi
=& \frac{1}{\zeta_0}\frac{\sin 2\theta}{8\pi}\left[
\sin\!\left(\delta_0-\delta_2\right)F_0(p)F_2(p)\right],\\
\zeta_z
=& \frac{1}{\zeta_0}\frac{1}{12\pi} \Bigg[
\frac{-1+3\cos 2\theta}{2}F_2^2(p)+F_0^2(p) \\
& \quad+ \frac{1+3\cos 2\theta}{2}F_2(p)F_0(p)
\cos\!\left(\delta_0-\delta_2\right)
\Bigg].
\end{aligned}
\right.
\end{equation}
Here, \(F_0(p)\) and \(F_2(p)\) denote the radial dipole transition amplitudes for the outgoing \(s\)- and \(d\)-wave channels, respectively, while \(\delta_0\) and \(\delta_2\) are the corresponding scattering phases. The number of independent observables matches the number of unknown quantities, allowing an algebraic extraction of the relative phase \(\Delta\delta_l=\delta_0-\delta_2\); see the Supplemental Material for details \cite{supp}.

Figure~\ref{FigSPI}(a) shows the relative scattering phase \(\Delta\delta_l\) for the \(j=\tfrac{3}{2}\) and \(j=\tfrac{1}{2}\) channels, extracted from the PST induced by an attosecond pulse train~\cite{paul2001observation}, together with the exact values obtained directly from the scattering-state wave functions. The extracted phase agrees well with the exact result for both channels. The vertical dashed lines mark the discrete photoelectron energies corresponding to different harmonic orders, \(E=\omega-I_p^{(j)}\). 
The slight horizontal offset between the two channels arises from the spin--orbit splitting of the initial state. From the energy dependence of \(\Delta\delta_l\), we obtain the corresponding Wigner time-delay difference \cite{wigner1955lower,smith1960lifetime},
\begin{equation}
\Delta t_W=\frac{\partial \Delta\delta_l}{\partial E}.
\end{equation}
As shown in Fig.~\ref{FigSPI}(b), the reconstructed \(\Delta t_W\) also agrees reasonably well with the exact result for both channels. The remaining discrepancy, about \(10\%\), arises from the finite energy resolution of the extraction. These results demonstrate that the spin torus provides a direct route to relative time-delay metrology in the single-photon regime, free from continuum--continuum coupling.

\end{document}